\documentclass[12pt]{amsart}
\usepackage{amssymb}
\usepackage{color}
\usepackage{amsmath,epic,curves}
\usepackage[english]{babel}
\usepackage{graphicx}
\newtheorem{claim}{}[section]
\newtheorem{theorem}[claim]{Theorem}

\newtheorem{proposition}[claim]{Proposition}

\renewenvironment{proof}{\noindent{\it Proof. \hskip0pt}}
                      {$\square$\par\medskip}

\begin{document}
\baselineskip 6.0 truemm
\parindent 1.5 true pc

\newcommand\lan{\langle}
\newcommand\ran{\rangle}
\newcommand\tr{{\text{\rm Tr}}\,}
\newcommand\ot{\otimes}
\newcommand\ol{\overline}
\newcommand\join{\vee}
\newcommand\meet{\wedge}
\renewcommand\ker{{\text{\rm Ker}}\,}
\newcommand\im{{\text{\rm Im}}\,}
\newcommand\id{{\text{\rm id}}}
\newcommand\tp{{\text{\rm tp}}}
\newcommand\pr{\prime}
\newcommand\e{\epsilon}
\newcommand\la{\lambda}
\newcommand\inte{{\text{\rm int}}\,}
\newcommand\ttt{{\text{\rm t}}}
\newcommand\spa{{\text{\rm span}}\,}
\newcommand\conv{{\text{\rm conv}}\,}
\newcommand\rank{\ {\text{\rm rank of}}\ }
\newcommand\re{{\text{\rm Re}}\,}
\newcommand\ppt{\mathbb T}
\newcommand\rk{{\text{\rm rank}}\,}

\title{Geometry of the faces for separable states\\ arising from generalized Choi maps}

\author{Kil-Chan Ha}
\address{Faculty of Mathematics and Statistics, Sejong University, Seoul 143-747, Korea}

\author{Seung-Hyeok Kye}
\address{Department of Mathematics and Institute of Mathematics\\Seoul National University\\Seoul 151-742, Korea}

\thanks{KCH is partially supported by NRFK 2011-0006561. SHK is partially supported by NRFK 2012-0000939}



\begin{abstract}
We exhibit examples of separable states which are on the boundary of the convex cone generated by all separable states but in the interior of the convex cone generated by all PPT states. We also analyze the geometric structures of the smallest face generated by those examples. As a byproduct, we obtain a large class of entangled states with positive partial transposes.
\end{abstract}

\maketitle

\section{Introduction}

The notion of entanglement is now one of the key research topics of quantum physics and considered as the main resources for quantum information and quantum computation theory. One of the most basic research topics in the theory of entanglement is how to distinguish entanglement from separable states. We recall that a state on the
tensor product $M_m\otimes M_n$ is said to be separable if it is the convex combination of product states, where
$M_n$ denotes the $*$-algebra of all $n\times n$ matrices over the complex field. If we identify a state on $M_m\otimes M_n$ as a density matrix in $M_m\otimes M_n$, then a density matrix is separable if and only if
it is the convex sum of rank one projections onto product vectors. A density matrix in $M_m\otimes M_n$ is entangled if it is not separable. 
It was observed by Choi \cite{choi-ppt} and Peres \cite{peres} that the partial transpose of a separable density matrix is again positive semi-definite. This PPT criterion gives us a very simple necessary condition for separability among various criteria for separability. See \cite{BZ}. The notion of PPT states turns out be to very important in itself in relation with bound entanglement. See \cite{mpr_horo}.

In order to distinguish entanglement from separability, it is important to understand the boundary structures of the convex cone $\mathbb V_1$ generated by all separable states. We note that the boundary $\partial\mathbb V_1$ of the cone $\mathbb V_1$ consists of maximal faces. We also note that the PPT criterion tells us that the cone $\mathbb V_1$ is contained in the convex cone $\mathbb T$ generated by all PPT states. Therefore, we have the two cases for a maximal face $F$ of $\mathbb V_1$:
\begin{itemize}
\item
$F$ is contained in the boundary $\partial\mathbb T$ of the cone $\mathbb T$,
\item
The interior ${\text{\rm int}}\, F$ of $F$ is contained in ${\text{\rm int}}\,\mathbb T$.
\end{itemize}
Although the first case is easy to characterize using the boundary structures \cite{ha_kye_04} of the cone $\mathbb T$,
there is nothing to be known for the second case. Even, there is no known explicit examples of separable states on the boundary
$\partial\mathbb V_1$ of $\mathbb V_1$, but in the interior of the cone $\mathbb T$, to the best knowledge of the authors.
The main purpose of this note is to exhibit such examples in the case of $M_3\otimes M_3$, and
investigate the geometric structures of the maximal faces determined by those examples. As a byproduct of those examples,
we naturally get examples of entangled states with positive partial transposes whose ranges are full spaces.

Note that every point $x$ of a convex set $C$ determines the smallest face containing the point.
This is the unique face of $C$ in which $x$ is an interior point.
In the next section,
we provide a general method how to understand the face of the cone $\mathbb V_1$ determined by a given separable state.
After we give examples of separable states in $\partial\mathbb V_1\cap{\text{\rm int}}\, \mathbb T$ in Section 3, we analyze
the faces determined by them in the last two sections. Throughout this note, a vector $z$ will be considered as a column vector, and we denote by $\bar z$ and $z^*$ the complex conjugate and the Hermitian conjugate of $z$, respectively. So
$zz^*$ is the one dimensional projection onto the vector $z$ with this notation, whenever $z$ is a unit vector.

\section{Faces for separable states}

The convex cone $\mathbb T$ consists of positive semi-definite matrices $A=\sum e_{ij}\otimes x_{ij}$ in $M_m\otimes M_n$ whose partial
transpose $A^\Gamma=\sum e_{ij}\otimes x_{ji}$ is also positive semi-definite, where $\{e_{ij}\}$ is the usual matrix units in $M_m$.
We recall that
a convex subset $F$ of a convex set $C$ is said to be a face of $C$ if the following property holds:
$$
x,\,y\in C,\ (1-t)x+ty \in F \ \text{for some } t\in (0,1)\ \Longrightarrow\ x,\,y\in F.
$$
In short, if an interior point of a line segment lies in the face $F$ then the whole line segment should lie in $F$.
The faces of the convex cone $\mathbb T$ are determined \cite{ha_kye_04}
by pairs $(D,E)$ of subspaces of $\mathbb C^m\otimes\mathbb C^n$. More precisely, every face of $\mathbb T$ is of the form
$$
\tau(D,E):=\{A\in\mathbb T: {\mathcal R}A\subset D,\ {\mathcal R}A^\Gamma\subset E\},
$$
where ${\mathcal R}A$ denotes the range space of $A\in M_m\otimes M_n$.
We also note that the interior of the face $\tau(D,E)$ is given by
\begin{equation}\label{eq:int_cond}
{\text{\rm int}}\, \tau(D,E)=\{A\in\mathbb T: {\mathcal R}A= D,\ {\mathcal R}A^\Gamma= E\}.
\end{equation}
We recall that a point $x$ of a convex set $C$ is an interior point of $C$ if the line segment from any point of $C$ to $x$
can be extended within $C$. A point of $C$ which is not an interior point is said to be a boundary point.
We denote by ${\text{\rm int}}\, C$ the set of all interior point of $C$, which is nothing but
the relative topological interior of $C$ with respect to the affine manifold generated by $C$.
The relation~\eqref{eq:int_cond} tells us that a separable state $A\in\mathbb V_1$ lies in the interior
of the convex cone $\mathbb T$ if and only if both ${\mathcal R}A$ and ${\mathcal R}A^\Gamma$ are the full space $\mathbb C^m\otimes\mathbb C^n$.
We also note that if $A\in\mathbb T$ then $\tau({\mathcal R}A,{\mathcal R}A^\Gamma)$ is the smallest face of $\mathbb T$
determined  by $A$, which will be denoted by $\mathbb T[A]$. That is, we define
$$
\mathbb T[A]:=\tau({\mathcal R}A,{\mathcal R}A^\Gamma).
$$

For a separable state $A\in\mathbb V_1$, we denote by $\mathbb V_1[A]$ the smallest face of $\mathbb V_1$ determined by $A$, and
define the following two sets of product vectors:
$$
\begin{aligned}
P[A]&=\{z=x\otimes y: zz^*\in \mathbb V_1[A]\},\\
Q[A]&=\{z=x\otimes y: x\otimes y\in{\mathcal R}A,\ \bar x\otimes y\in{\mathcal R}A^\Gamma\}.
\end{aligned}
$$
We note that $P[A]$ represents the set of all extreme rays of the face $\mathbb V_1[A]$.
For a set $P$ of product vectors, we denote by $\overline P$ the set of partial conjugates $\bar x\otimes y$
of $x\otimes y\in P$.
If two separable states $A,B\in\mathbb V_1$ lie in the interior of the same face $\tau(D,E)$ of $\mathbb T$,
then $Q[A]$ coincides with $Q[B]$. But, $P[A]$ and $P[B]$ may be different, even though they are in the interior
of a common face of $\mathbb T$.

\begin{theorem}{Proposition}
For every separable state $A\in\mathbb V_1$, we have the following:
\begin{enumerate}
\item[{\rm(i)}]
$P[A]\subset Q[A]$.
\item[{\rm(ii)}]
${\text{\rm span}}\, P[A]={\text{\rm span}}\, Q[A]={\mathcal R}A$.
\item[{\rm(iii)}]
${\text{\rm span}}\, \overline{P[A]}={\text{\rm span}}\, Q[A^\Gamma]={\mathcal R}A^\Gamma$.
\end{enumerate}
\end{theorem}

\begin{proof}
Suppose that $A$ is expressed by
\begin{equation}\label{22211}
A=\sum_\iota (x_\iota\otimes y_\iota)(x_\iota\otimes y_\iota)^*,
\end{equation}
for $x_\iota\otimes y_\iota\in\mathbb C^m\otimes\mathbb C^n$.
Then $A^\Gamma=\sum_\iota (\bar x_\iota\otimes y_\iota)(\bar x_\iota\otimes y_\iota)^*$, and we have
\begin{equation}\label{1111}
{\mathcal R}A={\text{\rm span}}\,\{x_\iota\otimes y_\iota\},\qquad {\mathcal R}A^\Gamma={\text{\rm span}}\,\{\bar x_\iota\otimes y_\iota\},
\end{equation}
by \cite{hugh}.
Let $z=x\otimes y$ be a product vector in $P[A]$, and so $zz^*\in \mathbb V_1[A]$.
Since $A$ is an interior point of $\mathbb V_1[A]$, we see that $A$ can be written
as the sum of $zz^*$ and a separable state $B=\sum z_\iota z_\iota^*\in \mathbb V_1[A]$ with product vectors $z_\iota$.
Therefore, we see that $x\otimes y\in{\mathcal R}A$,
and $\bar x\otimes y\in{\mathcal R}A^\Gamma$. This completes the proof of (i).

Since $x\otimes y\in Q[A]$ implies $x\otimes y\in{\mathcal R}A$ by definition, it is clear that ${\text{\rm span}}\, Q[A]\subset {\mathcal R}A$.
On the other hand, the expression (\ref{22211}) tells us that $(x_\iota\otimes y_\iota)(x_\iota\otimes y_\iota)^*$ belongs to the face
$\mathbb V_1[A]$, and so $x_\iota\otimes y_\iota\in P[A]$. By the relation (\ref{1111}), we have ${\mathcal R}A\subset {\text{\rm span}}\, P[A]$.
Similarly, we also have ${\mathcal R}A^\Gamma\subset {\text{\rm span}}\,\overline{P[A]}$.
\end{proof}

Now, we proceed to compare the face $\mathbb V_1[A]$ of the cone
$\mathbb V_1$ and the face $\mathbb T[A]$ of the cone $\mathbb T$
determined by $A\in\mathbb V_1\subset\mathbb T$. Since $A$ is an
interior point of both of $\mathbb V_1[A]$ and $\mathbb T[A]$, we see that
$\mathbb V_1[A]$ lies inside of $\mathbb T[A]$. More precisely, we
have
$$
{\text{\rm int}}\, \mathbb V_1[A]\subset {\text{\rm int}}\,\mathbb T[A],
\qquad
\mathbb V_1[A]\subset \mathbb V_1\cap \mathbb T[A].
$$
Since both convex set $\mathbb V_1[A]$ and $\mathbb V_1\cap \mathbb T[A]$ are faces of the convex cone $\mathbb V_1$,
we have the following two possibilities:
\begin{itemize}
\item
$\mathbb V_1[A]=\mathbb V_1\cap \mathbb T[A]$,
\item
$\mathbb V_1[A]$ is a proper face of $\mathbb V_1\cap \mathbb T[A]$.
\end{itemize}
If the first case occurs, then we see that $\mathbb V_1[A]$ is induced by
the face $\mathbb T[A]$ of the cone $\mathbb T$ in the sense of \cite{choi_kye}.
If both ${\mathcal R}A$ and ${\mathcal R}A^\Gamma$ are the full space $\mathbb C^m\otimes\mathbb C^n$ then the second case gives rise to
the boundary of the cone $\mathbb V_1$ which lies in the interior of the cone $\mathbb T$.

\begin{theorem}{Theorem}\label{the}
For a separable state $A\in\mathbb V_1$, the following are equivalent:
\begin{enumerate}
\item[{\rm (i)}]
$\mathbb V_1[A]=\mathbb V_1\cap \mathbb T[A]$
\item[{\rm (ii)}]
$A$ is an interior point of $\mathbb V_1\cap \mathbb T[A]$.
\item[{\rm (iii)}]
$P[A]=Q[A]$.
\item[{\rm (iv)}]
For every $z\in Q[A]$ there exist product vectors $z_\iota\in Q(A)$ such that $A=zz^*+\sum_\iota z_\iota z_\iota^*$.
\end{enumerate}
\end{theorem}

\begin{proof}
The equivalence (i) $\Longleftrightarrow$ (ii) follows from the definition of $\mathbb V_1[A]$.
We suppose that the condition (i) holds. Then we have
$$
x\otimes y\in P[A]
\ \Longleftrightarrow\
(x\otimes y)(x\otimes y)^*\in\mathbb T[A]
\ \Longleftrightarrow\
x\otimes y\in {\mathcal R}A,\ \bar x\otimes y\in {\mathcal R}A^\Gamma,
$$
since $\mathbb T[A]=\tau({\mathcal R}A,{\mathcal R}A^\Gamma)$.
This proves the direction (i) $\Longrightarrow$ (iii).
To prove the direction (iii) $\Longrightarrow$ (i),
suppose that $P[A]= Q[A]$. It suffices to show that $\mathbb V_1[A]\supset\mathbb V_1\cap \tau({\mathcal R}A,{\mathcal R}A^\Gamma)$.
To do this, take arbitrary $B=\sum z_\iota z_\iota^*\in \tau({\mathcal R}A,{\mathcal R}A^\Gamma)$
with $z_\iota=x_\iota\otimes y_\iota$.
Then we see that $x_\iota\otimes y_\iota\in \mathcal R A$ and $\bar x_\iota\otimes y_\iota \in \mathcal R A^{\tau}$,
which implies that $z_\iota\in Q[A]=P[A]$. This means that $z_\iota z_\iota^*\in \mathbb V_1[A]$, and so
we have $B\in \mathbb V_1[A]$.

To complete the proof, it remains to show (ii) $\Longleftrightarrow$ (iv). To do this, we first note that
the set of all extreme rays of the convex cone $\mathbb V_1\cap \mathbb T[A]$
consists of $zz^*$ with $z\in Q[A]$, since $zz^*\in \mathbb T[A]$ if and only if $z\in Q[A]$.
The statement (iv) says that the line segment from $zz^*$ to $A$ can be extended within
$\mathbb V_1\cap \mathbb T[A]$. Since every element in
$\mathbb V_1\cap \mathbb T[A]$ is the convex combination of $zz^*$ with $z\in Q[A]$,
we see that this is equivalent to say that
$A$ is an interior point of $\mathbb V_1\cap \mathbb T[A]$.
\end{proof}

When we consider a specific example of a separable state $A$, it is not so easy to determine the set $P[A]$
which characterize the extremal rays of the face $\mathbb V_1[A]$.
We note that it is relatively easy to determine the set $Q[A]$, and so we may determine
whether the condition (iv) of Theorem \ref{the} holds or not.
If the condition (iv) holds, then we may figure out the face $\mathbb V_1[A]$ in two ways: It is induced by
a face of the bigger cone $\mathbb T$ by (i) of Theorem \ref{the}; we know all extremal rays of the face
$\mathbb V_1[A]$ by (iii) of Theorem \ref{the}.

Another way to determine the set $P[A]$ is to use the duality \cite{eom-kye,horo-1} between the convex cone $\mathbb V_1$ and the cone
$\mathbb P_1$ consisting of all positive linear maps from $M_m$ into $M_n$.
For $A=\sum_{i,j=1}^m e_{ij}\otimes a_{ij}\in M_m\otimes M_n$ and a linear map $\phi$ from $M_m$ into $M_n$, we define the bilinear pairing by
\[
\langle A,\phi\rangle =\sum_{i,j=1}^m \text{Tr}(\phi(e_{ij})a_{ij}^{\rm t})=\text{Tr}(C_{\phi}A^{\rm t}),
\]
where
$C_{\phi}=\sum_{i,j=1}^m e_{ij}\otimes \phi(e_{ij})$
is the Choi matrix of the linear map $\phi$.
We note that every positive linear map $\phi$ gives rise to the face
$$
\phi'=\{A\in \mathbb V_1\,:\, \langle A,\phi\rangle=0\}
$$
of $\mathbb V_1$. We call $\phi'$ the dual face of the map $\phi$ with respect to dual pair $(\mathbb V_1,\mathbb P_1)$.
A face is said to be exposed if it is a dual face. It is not known that if every face of the cone $\mathbb V_1$ is exposed,
even though it is known \cite{ha_kye_04} that every face of the cone $\mathbb T$ is exposed. We refer to \cite{kye_ritsu}
for general aspects of duality.

For a product vector $z=x\otimes y\in\mathbb C^m\otimes\mathbb C^n$, we have
\begin{equation}\label{1-simple}
\langle zz^*,\phi\rangle
=\langle xx^*\otimes yy^*,\phi\rangle
={\text{\rm Tr}}\,(\phi(xx^*)\bar y\bar y^*)
=(\phi(xx^*)\bar y|\bar y),
\end{equation}
where $(\,\cdot\, |\,\cdot\,)$ denotes the inner product in $\mathbb C^n$ which is linear in the first variable and
conjugate linear in the second variable.
Therefore, we have
$z=x\otimes y\in P[A]$ if and only if $\langle zz^*,\phi\rangle=0$ if and only $\phi(xx^*)\in M_n$ is singular and $\bar y$ is
a kernel vector of $\phi(xx^*)$. In this way, we can determine the set $P[A]$.
See \cite{kye-canad} for related topics.

\section{Examples}

Recall  the generalized Choi map $\Phi[\alpha,\beta,\gamma]$ between $M_3$ defined by
$$
\begin{aligned}
&\Phi[\alpha,\beta,\gamma](X)\\
=&\begin{pmatrix}
\alpha x_{11}+\beta x_{22}+\gamma x_{33} & -x_{12} & -x_{13} \\
-x_{21} & \gamma x_{11}+\alpha x_{22}+\beta x_{33} & -x_{23} \\
-x_{31} & -x_{32} & \beta x_{11}+\gamma x_{22}+\alpha x_{33}
\end{pmatrix},
\end{aligned}
$$
for $X\in M_3$ and nonnegative real numbers $\alpha,\beta$ and $\gamma$, as was introduced in \cite{cho-kye-lee}.
It is a positive linear map if and only if the condition
$$
\alpha+\beta+\gamma\ge 2,\qquad 0\le \alpha\le 1\ \Longrightarrow\ \beta\gamma\ge (1-\alpha)^2
$$
holds. The Choi matrix $C_\Phi[\alpha,\beta,\gamma]$ of the map $\Phi[\alpha,\beta,\gamma]$ is given by
$$
C_\Phi[\alpha,\beta,\gamma]=\left(
\begin{array}{ccccccccccc}
\alpha     &\cdot   &\cdot  &\cdot  &-1     &\cdot   &\cdot   &\cdot  &-1     \\
\cdot   &\gamma &\cdot    &\cdot    &\cdot   &\cdot &\cdot &\cdot     &\cdot   \\
\cdot  &\cdot    &\beta &\cdot &\cdot  &\cdot    &\cdot    &\cdot &\cdot  \\
\cdot  &\cdot    &\cdot &\beta &\cdot  &\cdot    &\cdot    &\cdot &\cdot  \\
-1     &\cdot   &\cdot  &\cdot  &\alpha     &\cdot   &\cdot   &\cdot  &-1     \\
\cdot   &\cdot &\cdot    &\cdot    &\cdot   &\gamma &\cdot &\cdot    &\cdot   \\
\cdot   &\cdot &\cdot    &\cdot    &\cdot   &\cdot &\gamma &\cdot    &\cdot   \\
\cdot  &\cdot    &\cdot &\cdot &\cdot  &\cdot    &\cdot    &\beta &\cdot  \\
-1     &\cdot   &\cdot  &\cdot  &-1     &\cdot   &\cdot   &\cdot  &\alpha
\end{array}
\right)\in M_3\otimes M_3.
$$
The most interesting cases occur when the following condition
\begin{equation}\label{cond}
0\le \alpha< 1,\qquad \alpha+\beta+\gamma= 2,\qquad \beta\gamma= (1-\alpha)^2
\end{equation}
holds. Motivated by the parametrization in \cite{cw} for those cases,
the authors \cite{ha+kye_exposed} have shown that $\Phi[\alpha,\beta,\gamma]$ is an indecomposable exposed positive
linear map under the conditions (\ref{cond}). To do this, they considered the another parametrization given
by
$$
\Phi(t)= \Phi\left[\dfrac{(1-t)^2}{1-t+t^2},\ \dfrac {t^2}{1-t+t^2},
\dfrac 1{1-t+t^2}\right],\qquad 0<t<\infty.
$$
See also \cite{cs} and \cite{ha+kye_indec-witness}.
Note that $\Phi(1)=\Phi[0,1,1]$ is completely copositive, and
both $\Phi(0)=\Phi[1,0,1]$ and $\Phi(\infty)=\Phi[1,1,0]$ are indecomposable extremal positive maps \cite{choi-lam}
which are not exposed. See the picture in Section 5 of \cite{kye_ritsu}.

We note that the matrix $C_\Phi[\alpha,\beta,\gamma]$ is of PPT if and only if $\alpha\ge 2$ and $\beta\gamma\ge 1$.
It turns out \cite{kye_osaka} that it is in fact separable if $\alpha=2$ and $\beta\gamma=1$. Therefore, we see
that $C_\Phi[\alpha,\beta,\gamma]$ is of PPT if and only if it is separable. Our strategy is to consider
the following matrix
\begin{equation}\label{eq:form}
A[a,b,c]=\left(
\begin{array}{ccccccccccc}
a     &\cdot   &\cdot  &\cdot  &1     &\cdot   &\cdot   &\cdot  &1     \\
\cdot   &c &\cdot    &\cdot    &\cdot   &\cdot &\cdot &\cdot     &\cdot   \\
\cdot  &\cdot    &b &\cdot &\cdot  &\cdot    &\cdot    &\cdot &\cdot  \\
\cdot  &\cdot    &\cdot &b &\cdot  &\cdot    &\cdot    &\cdot &\cdot  \\
1     &\cdot   &\cdot  &\cdot  &a     &\cdot   &\cdot   &\cdot  &1     \\
\cdot   &\cdot &\cdot    &\cdot    &\cdot   &c &\cdot &\cdot    &\cdot   \\
\cdot   &\cdot &\cdot    &\cdot    &\cdot   &\cdot &c &\cdot    &\cdot   \\
\cdot  &\cdot    &\cdot &\cdot &\cdot  &\cdot    &\cdot    &b &\cdot  \\
1     &\cdot   &\cdot  &\cdot  &1     &\cdot   &\cdot   &\cdot  &a
\end{array}
\right)\in M_3\otimes M_3
\end{equation}
for nonnegative real numbers $a,b$ and $c$, and seek the condition for separability using the indecomposable
exposed positive linear map
$\Phi(t)$. Note that
$$
\langle A[a,b,c], \Phi[\alpha,\beta,\gamma]\rangle ={\text{\rm Tr}}\, (A[a,b,c]C_\Phi[\alpha,\beta,\gamma]^{\text{\rm t}})=
3(a\alpha+b\beta+c\gamma-2).
$$
First of all, it is easy to see that $A$ is of PPT if and only if
\begin{equation}\label{ppt}
a\ge 1,\qquad bc\ge 1,
\end{equation}
since we have
$$
A[a,b,c]^\Gamma=\left(
\begin{array}{ccccccccccc}
a     &\cdot   &\cdot  &\cdot  &\cdot     &\cdot   &\cdot   &\cdot  &\cdot     \\
\cdot   &c &\cdot    &1    &\cdot   &\cdot &\cdot &\cdot     &\cdot   \\
\cdot  &\cdot    &b &\cdot &\cdot  &\cdot    &1    &\cdot &\cdot  \\
\cdot  &1    &\cdot &b &\cdot  &\cdot    &\cdot    &\cdot &\cdot  \\
\cdot     &\cdot   &\cdot  &\cdot  &a     &\cdot   &\cdot   &\cdot  &\cdot     \\
\cdot   &\cdot &\cdot    &\cdot    &\cdot   &c &\cdot &1    &\cdot   \\
\cdot   &\cdot &1    &\cdot    &\cdot   &\cdot &c &\cdot    &\cdot   \\
\cdot  &\cdot    &\cdot &\cdot &\cdot  &1    &\cdot    &b &\cdot  \\
\cdot     &\cdot   &\cdot  &\cdot  &\cdot     &\cdot   &\cdot &\cdot  &a
\end{array}
\right).
$$
A PPT state $A$ is said to be with type $(p,q)$ if $\dim{\mathcal R}A=p$ and $\dim{\mathcal R}A^\Gamma=q$.
We note that $A[a,b,c]$ is of PPT with type
\begin{itemize}
\item
$(7,6)$ if $a=1$ and $bc=1$,
\item
$(9,6)$ if $a>1$ and $bc=1$,
\item
$(7,9)$ if $a=1$ and $bc>1$,
\item
$(9,9)$ if $a>1$ and $bc>1$.
\end{itemize}

To get a necessary condition for separability of $A[a,b,c]$, we suppose that $A[a,b,c]$ is separable. Then
we have $\langle A[a,b,c], \Phi(t)\rangle \ge 0$ for each $t>0$. In other words, we have the condition
$$
t >0\ \Longrightarrow\ \dfrac{a(1-t)^2}{1-t+t^2}+\dfrac{bt^2}{1-t+t^2}+\dfrac{c}{1-t+t^2}\ge 2.
$$
This condition holds if and only if
$$
t >0\ \Longrightarrow\ a(1-t)^2+bt^2+c\ge 2(1-t+t^2)
$$
if and only if
$$
t >0\ \Longrightarrow\ (a+b-2)t^2+2(1-a)t+(a+c-2)\ge 0
$$
if and only if
\begin{equation}\label{sep}
a+b-2\ge 0,\qquad (b+a-2)(c+a-2)\ge (1-a)^2.
\end{equation}
Note that \eqref{sep} implies $a+c-2\ge 0$, which may replace $a+b-2\ge 0$ in \eqref{sep}.
Therefore, we see that if $A[a,b,c]$ is separable then the both conditions (\ref{ppt}) and (\ref{sep}) hold.
We proceed to show that these are sufficient for separability of $A[a,b,c]$. We note that the condition
(\ref{sep}) implies (\ref{ppt}) strictly when $1\le a< 2$, and two conditions (\ref{ppt}) and (\ref{sep}) coincide
when $a=2$.

We denote by $C$ the convex subset consisting of $(a,b,c)\in \mathbb R^3$ satisfying the conditions (\ref{ppt}) and (\ref{sep}).
We see that the following three points
$$
(1,1,1),\qquad (a,y,z),\qquad \left(2,\dfrac{y-1}{a-1}+1,\dfrac{z-1}{a-1}+1\right)
$$
are on a single line for $1<a<2$, and if the triplet $(a,b,c)=(a,y,z)$ satisfies the equality in the second inequality of (\ref{sep}) then we have
$$
\left(\dfrac{y-1}{a-1}+1\right)\left(\dfrac{z-1}{a-1}+1\right)=1.
$$
Therefore, if we put $b=\frac{y-1}{a-1}+1$ and $c=\frac{z-1}{a-1}+1$ then the point $(2,b,c)$ lies on the boundary of the
convex body determined by (\ref{ppt}) and (\ref{sep}).
This means that every boundary point of the convex body
$$
C_1=\{(a,y,z): 1<a<2,\ a,y,z\ {\text{\rm satisfy (\ref{sep})}}\}
$$
lies on the line segment between the point $(1,1,1)$ and a point $(2,b,\frac1b)$ for $b\neq 1$, which are extreme points of the convex body
$C$.
In conclusion, the boundary of the convex body $C$ consists of the following:
\begin{figure}[h!]
\begin{center}
\includegraphics[scale=0.8]{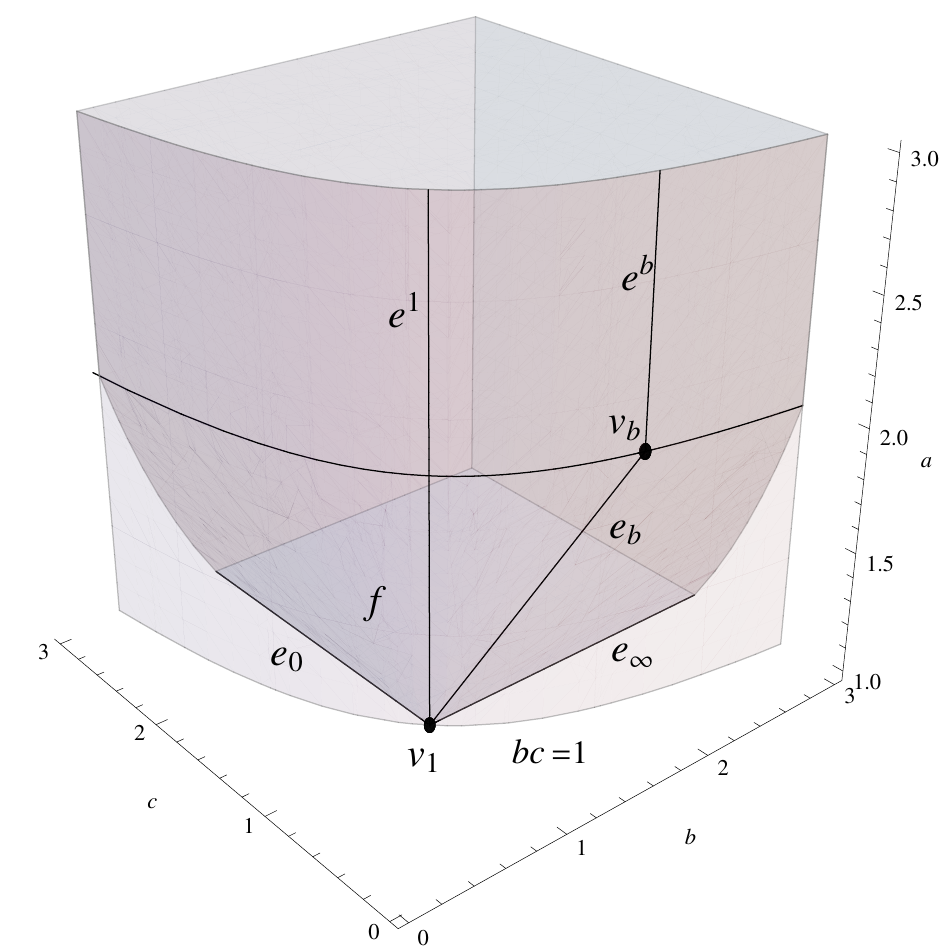}
\end{center}
\caption{The edge $e_{\infty}$ can be regarded as the limit of $e_b$ as $b\to\infty$, and $e_0$ can be regarded as the limit of $e_b$ as $b\to0$. }
\end{figure}
\begin{itemize}
\item
$v_1=(1,1,1)$
\item
$v_b=(2,b,\frac 1b)$ for $b>0$ with $b\neq 1$
\item
$e^1=\{(a,1,1):a\ge 1\}$
\item
$e^b=\{(a,b,\frac 1b): a\ge 2\}$  for $b>0$ with $b\neq 1$
\item
$e_b=$ the line segment between $v_1$ and $v_b$  for $b>0$ with $b\neq 1$
\item
$e_0=\{(1,1,c): c\ge 1\}$
\item
$e_\infty=\{(1,b,1): b\ge 1\}$
\item
$f=\{(1,b,c): b\ge 1,\ c\ge 1\}$
\end{itemize}
Any interior points of edges $e^1,e^b,e_b,e_0,e_\infty$ and two-dimensional face $f$ will be denoted by
the same symbols. For example, any point $(a,b,\frac 1b)$ with $a>2$ will be denoted  by just $e^b$. This does not
make any confusion, because any two interior points of the \lq edge\rq\ $e_b$ determine the same face
as that determined by the edge $e_b$ itself.

In order to prove that the condition (\ref{sep}) together with (\ref{ppt}) implies the separability
of $A[a,b,c]$, it suffices to show that
$A[v_b]$ are separable for each $b>0$.
We first consider the case of $b\neq 1$. To do this, we define product vectors
\begin{equation}\label{eq:z_def}
\begin{aligned}
z_1(\omega)&=(0,1,\sqrt b\,\omega)^{\text{\rm t}}\otimes (0,\sqrt b,\bar\omega)^{\text{\rm t}}
=(0,0,0 \,;\, 0,\sqrt b, \bar\omega \,;\, 0,b\, \omega,\sqrt b\,)^{\text{\rm t}},\\
z_2(\omega)&=(\sqrt b\,\omega,0,1)^{\text{\rm t}}\otimes (\bar\omega,0,\sqrt b)^{\text{\rm t}}
=(\sqrt b,0,b\, \omega \,;\, 0,0,0 \,;\, \bar\omega,0,\sqrt b\,)^{\text{\rm t}},\\
z_3(\omega)&=(1,\sqrt b\,\omega,0)^{\text{\rm t}}\otimes (\sqrt b,\bar\omega,0)^{\text{\rm t}}
=(\sqrt b,\bar\omega,0 \,;\, b\, \omega,\sqrt b,0 \,;\, 0,0,0)^{\text{\rm t}}
\end{aligned}
\end{equation}
in $\mathbb C^3\otimes\mathbb C^3$, for $\omega\in\mathbb C$ with $|\omega|=1$. Then it is straightforward to see that
$$
A[ v_b]=\dfrac 1{3b}\sum_{i=1}^3\sum_{\omega\in\Omega}z_i(\omega)z_i(\omega)^*,
$$
where $\Omega=\{1,e^{\frac 23\pi i}, e^{-\frac 23\pi i}\}$ is the third roots of unity.
If we put
\begin{equation}\label{eq:z_def_4}
z_4(\omega,\eta)=(1,\bar\omega,\bar{\eta})\otimes (1,\omega,\eta)
=(1,\omega,\eta\,;\, \bar\omega,1,\bar\omega\eta\,;\, \bar\eta, \omega\bar\eta,1)
\end{equation}
for $(\omega,\eta)\in\Omega\times\Omega$, then we also have
\begin{equation}\label{eq:A111}
A[v_1]=\dfrac 19\sum_{(\omega,\eta)\in\Omega\times\Omega}
z_4(\omega,\eta)z_4(\omega,\eta)^*.
\end{equation}
Therefore, we see that $A[v_1]$ and $A[v_b]$ are separable.

\begin{theorem}{Theorem}\label{thm:sep}
The state $A[a,b,c]$ is of PPT if and only if the condition {\rm (\ref{ppt})} holds,
and separable if and only if both conditions {\rm (\ref{ppt})} and {\rm (\ref{sep})} hold.
\end{theorem}

Now, we can characterize a large class of entangled states of the form
\eqref{eq:form} with positive partial transpose.
\begin{theorem}{Corollary}
The state $A[a,b,c]$ is PPTES if and only if the  condition
\[
1\le a<2,\quad bc\ge 1,\quad (b+a-2)(c+a-2)<(1-a)^2
\]
holds.
\end{theorem}

The state $A[1,b,\frac 1b]$ is nothing but the PPTES considered in \cite{stormer82} in the early eighties, which has been
reconstructed in \cite{ha+kye} systematically using the indecomposable positive linear maps.

\section{Maximal Faces}

In this section, we determine the faces $\mathbb V_1[A]$ for separable state $A$ which comes from the boundaries
of the convex body $C$.
We first determine the sets $Q[A[v_1]]$ and $Q[A[v_b]]$.
We note that the kernel of $A[v_1]$ is spanned by
$$
(1,0,0\,;\, 0,-1,0 \,;\, 0,0,0)^{\text{\rm t}},\qquad
(0,0,0\,;\, 0,1,0 \,;\, 0,0,-1)^{\text{\rm t}},
$$
and the kernel of $A[v_1]^\Gamma$ is spanned by
$$
\begin{aligned}
(0,1,0\,;\,-1,0,0\,;\,0,0,0)^{\text{\rm t}},\\
(0,0,0\,;\,0,0,1\,;\,0,-1,0)^{\text{\rm t}},\\
(0,0,-1\,;\,0,0,0\,;\,1,0,0)^{\text{\rm t}}.
\end{aligned}
$$
If $x\otimes y\in {\mathcal R}A[v_1]$ and $\bar x\otimes y\in {\mathcal R}A[v_1]^\Gamma$,
then $\bar x$ and $y$ are parallel to each other, and so we may
assume that $\bar x=y$. Further, we have $|x_1|=|x_2|=|x_3|$. Therefore, we see that
\begin{equation}\label{111}
Q[A[v_1]]=\{(\bar x_1,\bar x_2,\bar x_3)^{\text{\rm t}}\otimes (x_1,x_2,x_3)^{\text{\rm t}}:|x_1|=|x_2|=|x_3|\}.
\end{equation}

Next, we turn our attention to $A[v_b]$ for $b\neq 1$.
The kernel of $A[v_b]^\Gamma$ is spanned by
$$
\begin{aligned}
(0,b,0\,;\,-1,0,0\,;\,0,0,0)^{\text{\rm t}},\\
(0,0,0;0,0,b\,;\,0,-1,0)^{\text{\rm t}},\\
(0,0,-1\,;\,0,0,0\,;\,b,0,0)^{\text{\rm t}}
\end{aligned}
$$
If $\bar x\otimes y$ is a range vector of $A[v_b]^\Gamma$ then we have
\begin{equation}\label{2}
b\bar x_1 y_2-\bar x_2 y_1=0,\quad
b\bar x_2 y_3-\bar x_3 y_2=0,\quad
b\bar x_3 y_1-\bar x_1 y_3=0.
\end{equation}
Multiplying the above equations, we have $b^3\bar x_1\bar x_2\bar x_3 y_1 y_2 y_3=\bar x_1\bar x_2\bar x_3 y_1 y_2 y_3$, from which
we see that at least one of $x_i$ is zero.
We also have
$ x_i=0 \Longleftrightarrow  y_i=0$ from (\ref{2}).
Therefore, we see that the set $Q[A[v_b]]$ consists of product vectors
\begin{equation}\label{11/bb1}
\begin{aligned}
&(0,\bar x_2, b\bar x_3)^{\rm t}\otimes (0,x_2,x_3)^{\rm t},\\
&(b\bar x_1,0, \bar x_3)^{\rm t}\otimes (x_1,0,x_3)^{\rm t},\\
&(\bar x_1,b\bar x_2, 0)^{\rm t}\otimes (x_1,x_2,0)^{\rm t}.
\end{aligned}
\end{equation}

We proceed to show that $A[v_1]$ satisfies the condition (iv) of Theorem \ref{the}.
We take arbitrary $z\in Q[A[v_1]]$, then we may assume that
$$
z=(1,\bar\alpha,\bar\beta)^{\text{\rm t}}\otimes (1,\alpha,\beta)^{\text{\rm t}},
$$
with $|\alpha|=|\beta|=1$. Then we have the relation
$$
A[v_1]=\dfrac 19\sum_{(\omega,\eta)\in\Omega\times\Omega}
z_4(\alpha\omega,\beta\eta)z_4(\alpha\omega,\beta\eta)^*,
$$
where $z_4(\,\cdot\, ,\, \cdot\, )$ was defined in (\ref{eq:z_def_4}),
and so we see that $A[v_1]$ satisfies the conditions in Theorem \ref{the}.

We note that the correspondence $x\mapsto A[x]$ is an injective affine map from the convex body $C$ into the convex cone
$\mathbb V_1$, and the line segment from an interior point of the edge $e_b$ to the vertex $v_b$ cannot be
extended within $C$ whenever $b\neq 1$. This tells us that the
line segment from $A[e_b]$ to $A[v_b]$ cannot be extended within the cone $\mathbb V_1$.
We also note that ${\mathcal R}A[e_b]={\mathcal R}A[v_b]$ and ${\mathcal R}A[e_b]^\Gamma={\mathcal R}A[v_b]^\Gamma$,
and so $\mathbb T[A[e_b]]=\mathbb T[A[v_b]]$. Therefore, the line segment from $A[e_b]\in\mathbb V_1\cap \mathbb T[A[v_b]]$ to
$A[v_b]$ cannot be extended within the face $\mathbb V_1\cap \mathbb T[A[v_b]]$.
This means that that
$A[v_b]$ is on the boundary point of the face $\mathbb V_1\cap \mathbb T[A[v_b]]$, and so
$A[v_b]$ does not satisfy the conditions in Theorem \ref{the}.
We use the duality to overcome this difficulty.
We see by a direct calculation
$$
\langle A[v_1],\Phi(\textstyle\frac 1b)\rangle=\langle A[v_b],\Phi(\textstyle\frac 1b)\rangle=0
$$
that $A[v_1]$ and $A[v_b]$ belong to the dual face $\Phi(\frac 1b)^\prime$ of the map
$\Phi(\frac 1b)\in\mathbb P_1$ with respect to dual pair $(\mathbb V_1,\mathbb P_1)$.

Now, we determine the set $\{z\in Q[A[v_b]]: zz^*\in\Phi(\textstyle\frac 1b)^\prime\}$.
For $z=(\bar x_1,b\bar x_2, 0)^{\text{\rm t}}\otimes (x_1,x_2,0)^{\text{\rm t}}$, we see that $zz^*\in\Phi(\frac 1b)^\prime$ if and only if
$$
\begin{aligned}
|x_1|^4\dfrac{(1-b)^2}{1-b+b^2}
+&|x_1x_2|^2  \dfrac{b^2}{1-b+b^2}\\
&+b^2|x_1x_2|^2 \dfrac{1}{1-b+b^2}
+b^2|x_2|^4\dfrac{(1-b)^2}{1-b+b^2}-2b|x_1x_2|^2=0
\end{aligned}
$$
if and only if
$$
(|x_1|^2-b|x_2|^2)=0
$$
if and only if the relation $|x_1|^2=b|x_2|^2$ holds. By the similar way, we also see that
the set $\{z\in Q[A[v_b]]: zz^*\in\Phi(\textstyle\frac 1b)^\prime\}$ consists of product vectors
\begin{equation}\label{11/bb2}
\begin{aligned}
&(0,\bar x_2, b\bar x_3)^{\rm t}\otimes (0,x_2,x_3)^{\rm t}\quad {\text{\rm with}}\ |x_2|^2=b|x_3|^2,\\
&(b\bar x_1,0, \bar x_3)^{\rm t}\otimes (x_1,0,x_3)^{\rm t}\quad {\text{\rm with}}\ |x_3|^2=b|x_1|^2,\\
&(\bar x_1,b\bar x_2, 0)^{\rm t}\otimes (x_1,x_2,0)^{\rm t}\quad {\text{\rm with}}\ |x_1|^2=b|x_2|^2 ,
\end{aligned}
\end{equation}
as was done in \cite{ha+kye_exposed}.
We proceed to show that
$$
\mathbb V_1[A[v_b]]
=\mathbb T[A[v_b]]\cap \Phi(\textstyle\frac 1b)^\prime.
$$
To do this,
it suffices to show that $A[v_b]$ is an interior point of the convex set spanned by $zz^*$ with $z$ in (\ref{11/bb2}).
Take a $z$ in (\ref{11/bb2}), we may assume that $z$ is one of the following forms
$$
(0,1, b\bar \alpha)^{\rm t}\otimes (0,1,\alpha)^{\rm t},\quad
(b\bar \alpha,0,1)^{\rm t}\otimes (\alpha,0,1)^{\rm t},\quad
(1, b\bar \alpha,0)^{\rm t}\otimes (1,\alpha,0)^{\rm t},
$$
with $ b|\alpha|^2=1$. Note that we have
$$
\begin{aligned}
&(0,1, b\bar \alpha)^{\rm t}\otimes (0,1,\alpha)^{\rm t}=\dfrac 1{\sqrt b}z_1(\sqrt b \bar \alpha ),\\
&(b\bar \alpha,0,1)^{\rm t}\otimes (\alpha,0,1)^{\rm t}=\dfrac 1{\sqrt b}z_2(\sqrt b \bar \alpha ), \\
&(1, b\bar \alpha,0)^{\rm t}\otimes (1,\alpha,0)^{\rm t}=\dfrac 1{\sqrt b}z_3(\sqrt b \bar \alpha ),
\end{aligned}
$$
where $z_i(\cdot)$ is defined as in the equation~\eqref{eq:z_def}.
Therefore, the above claim is proved by the relation
$$
A[2,b,{\textstyle\frac 1b}]=\dfrac 1{3b}\sum_{i=1}^3\sum_{\omega\in\Omega}z_i(\sqrt{b} \bar\alpha \omega)\,z_i(\sqrt{b} \bar\alpha \omega)^*,
$$
where $\Omega=\{1,e^{\frac 23\pi i}, e^{-\frac 23\pi i}\}$ is again the third roots of unity.

Since $A[v_1]$ and $A[v_b]$ belong to the dual face $\Phi(\frac 1b)^\prime$, we see that the convex hull
generated by $A[v_1]$ and $A[v_b]$ is contained in the dual face $\Phi(\frac 1b)^\prime$.
For a product vector $z\in\mathbb C^3\otimes\mathbb C^3$,
we recall \cite{ha+kye_exposed} that $zz^*\in\Phi(\frac 1b)^\prime$ if and only if $z\in Q[A[v_1]]$ or $z$ is of the form in
(\ref{11/bb2}). This means that the face $\Phi(\frac 1b)^\prime$ is exactly the convex hull of two faces
$\mathbb V_1[A[v_1]]$ and $\mathbb V_1[A[v_b]]$.
Furthermore, since $A[v_1]$ and $A[v_b]$ are interior points of $\mathbb V_1[A[v_1]]$ and $\mathbb V_1[A[v_b]]$, respectively,
and $A[e_b]$ is a nontrivial convex combination of $A[v_1]$ and $A[v_b]$, we conclude that
$A[e_b]$ is an interior point of the convex hull of the faces $\mathbb V_1[A[v_1]]$ and $\mathbb V_1[A[v_b]]$.
We summarize as follows:

\begin{theorem}{Theorem}
We have the following:
\begin{enumerate}
\item[{\rm (i)}]
$\mathbb V_1[A[v_1]]=\mathbb V_1\cap \mathbb T[A[v_1]]$.
\item[{\rm (ii)}]
$\mathbb V_1[A[v_b]]=\Phi(\frac 1b)^\prime\cap
\mathbb T[A[v_b]]$, for every nonnegative real number $b$ with $b\neq 1$.
\item[{\rm (iii)}]
$\mathbb V_1[A[e_b]]=\Phi(\frac 1b)^\prime$ is the convex hull of $\mathbb V_1[A[v_1]]$ and $\mathbb V_1[A[v_b]]$,
for every nonnegative real number $b$ with $b\neq 1$.
\end{enumerate}
\end{theorem}

We note that both ${\mathcal R}A[e_b]$ and ${\mathcal R}A[e_b]^\Gamma$ are full spaces, and so
$\mathbb V_1[A[e_b]]$ gives rise to
an explicit example of proper face
of $\mathbb V_1$ which is in the interior of the cone $\mathbb T$.
Since $\Phi(\frac 1b)$ is exposed \cite{ha+kye_exposed}, we see that $\mathbb V_1[A[e_b]]$ is
a maximal faces of $\mathbb V_1$.

\section{Other Faces}

We have seen that $A[v_b]$ does not satisfy the conditions in Theorem \ref{the} for $b\neq 1$. The exactly same
argument shows that neither $A[v_0]$ nor $A[v_\infty]$ satisfy the conditions in Theorem \ref{the}, if we use
($f,e_0)$ and $(f,e_\infty)$ in the places of $(e_b,v_b)$ in the argument.
In the remaining of this paper, we show that
$A[e^1], A[e^b]$ and $A[f]$ satisfy the conditions in Theorem \ref{the}.
We will sustain the notation $\Omega$ for the set of third roots of unity.

First of all, we consider $A[e^b]$. Recall that $A[e^b]$ is an interior point of the face $\mathbb T[v_b]$.
Therefore, $Q[e^b]$ coincides with $Q[v_b]$, and product vectors which belong to $Q[e^b]$ are of the forms
 in \eqref{11/bb1}.
In order to show that $A[e^b]$ satisfies the condition (iv) of
Theorem \ref{the}, it suffices to consider the following product vector:
$$
\begin{aligned}
\widetilde{z_1}(\alpha)&=(0,1, b\,\bar\alpha)^{\text{\rm t}}\otimes (0,1,\alpha)^{\text{\rm t}}
=(0,0,0 \,;\, 0,1, \alpha \,;\, 0,b\, \bar\alpha,b\,|\alpha|^2\,)^{\text{\rm t}},\\
\widetilde{z_2}(\alpha)&=(b\,\bar\alpha,0,1)^{\text{\rm t}}\otimes (\alpha,0,1)^{\text{\rm t}}
=(b\,|\alpha|^2,0,b\, \bar\alpha \,;\, 0,0,0 \,;\,\alpha,0,1\,)^{\text{\rm t}},\\
\widetilde{z_3}(\alpha)&=(1, b\,\bar\alpha,0)^{\text{\rm t}}\otimes (1,\alpha,0)^{\text{\rm t}}
=(1,\alpha,0 \,;\, b\, \bar\alpha,b\,|\alpha|^2,0 \,;\, 0,0,0)^{\text{\rm t}}
\end{aligned}
$$
in $\mathbb C^3\otimes\mathbb C^3$, for $\alpha\in\mathbb C$. We fix an interior point $(a,b,\frac 1b)$
in the interior of the edge $e_b$ with $a>2$. It is straightforward to see that
$$
A[a,b,{\textstyle\frac1b}]=A[2,b,{\textstyle \frac1b}]+(a-2)\sum_{i=1}^3 \widetilde{z_i}(0) \widetilde{z_i}(0)^*.
$$
Now, we assume that $\alpha\neq 0$. Then we also have
$$
A[a(\alpha),b,{\textstyle \frac 1b}]
=\dfrac 1{3b\,|\alpha|^2}\sum_{i=1}^3\sum_{\omega\in\Omega}\widetilde{z_i}(\alpha \omega)\,\widetilde{z_i}(\alpha \omega)^*,
$$
where
$$
a(\alpha)=b\,|\alpha|^2+\dfrac1 {b\,|\alpha|^2}\ge 2.
$$
If $a(\alpha)=a$, then we have done. If $a(\alpha)\neq a$, then we can choose $t_0>1$ such that
$$
(1-t_0)a(\alpha)+t_0a>2.
$$
Therefore, we can take a complex number $\beta$ such that
$$
a(\beta)=(1-t_0)a(\alpha)+t_0a,
$$
and we have the relation
$$
A[a,b,{\textstyle\frac 1b}]=\dfrac 1{t_0} A[a(\beta),b,{\textstyle\frac1b}]+\left(1-\dfrac 1{t_0}\right) A[a(\alpha),b,{\textstyle\frac 1b}].
$$
Consequently, we conclude that $A[e^b]$ satisfies the conditions in Theorem~\ref{the}.

Next, we turn our attention to $A[e^1]$, that is, $A[a,1,1]$ for a fixed $a>1$.
First, it is easy to see that $Q[e^1]$ consists of product vectors
$$
\tilde z(x_1,x_2,x_3)= (x_1,x_2,x_3)^{\text{\rm t}}\otimes (\bar x_1,\bar x_2,\bar x_3)^{\text{\rm t}}.
$$
We take arbitrary $\tilde z(x_1,x_2,x_3)$ in $Q[e^1]$.  If the only one $x_i$ is nonzero in $\tilde z(x_1,x_2,x_3)$,  then we have the relation
$$
\begin{aligned}
A[a,&1,1]=A[1,1,1]\\ &+(a-1)\left (\tilde z(1,0,0) \tilde z(1,0,0)^*
+\tilde z(0,1,0) \tilde z(0,1,0)^*+\tilde z(0,0,1) \tilde z(0,0,1)^* \right).
\end{aligned}
$$
Now, we assume that at least two $x_i$'s are nonzero in $\tilde z(x_1,x_2,x_3)$, and define
$$
\begin{aligned}
k(x_1,x_2,x_3)&=|x_1|^2|x_2|^2+|x_2|^2|x_3|^2+|x_3|^2|x_1|^2,\\
a(x_1,x_2,x_3)&=\dfrac{|x_1|^4+|x_2|^4+|x_3|^4}{k(x_1,x_2,x_3)}.
\end{aligned}
$$
Then we get $a(x_1,x_2,x_3)\ge 1$, and it is starightforward to see that
$$
\begin{aligned}
A[a(x_1,& x_2,x_3),1,1]\\
=&\dfrac 1{9\, k(x_1,x_2,x_3)}\sum_{(s_1,s_2,s_3)\in \Lambda} \sum_{(\omega,\eta)\in\Omega\times\Omega}
\tilde z(s_1,\omega s_2,\eta s_3)\tilde z(s_1, \omega s_2,\eta s_3)^*,
\end{aligned}
$$
where $\Lambda =\{(x_1,x_2,x_3),\,(x_2,x_3,x_1),\,(x_3,x_1,x_2)\}$.
If $a(x_1,x_2,x_3)=a$, then we have done.
If $a(x_1,x_2,x_3)\neq a$, then we can choose $t_0>1$ and $\tilde x_1$ such that
$$
a(\tilde x_1,1,1)=(1-t_0) a(x_1,x_2,x_3)+t_0 a>1,
$$
and so we have the relation
$$
A[a,1,1]=\dfrac 1{t_0} A[a(\tilde x_1,1,1),1,1]+\left (1-\dfrac 1{t_0}\right) A[a(x_1,x_2,x_3),1,1].
$$
Consequently, we see that $A[e^1]$ satisfies the conditions of Theorem~\ref{the}.

Finally, we show that $A[f]$ satisfies the condition (iv) of Theorem~\ref{the}. We begin with description of $Q[A[f]]$. It is easy to see that
$$
Q[A[f]]=\{(x_1,x_2,x_3)^{\text{\rm t}}\,\otimes (y_1,y_2,y_3)^{\text{\rm t}}\,:\, x_1y_1=x_2y_2=x_3y_3\}.
$$
Therefore, we may assume that a product vector in $Q[A[f]]$ is one of the following vectors:
\begin{equation}\label{eq:vec_1bc}
\begin{aligned}
v_1(s,t)=&(1,0,0)^{\text{\rm t}}\otimes (0,s,t)^{\text{\rm t}}=(0,s,t,\,;\,0,0,0,\,;\,0,0,0)^{\text{\rm t}},\\
v_2(s,t)=&(0,1,0)^{\text{\rm t}}\otimes (s,0,t)^{\text{\rm t}}=(0,0,0,\,;\,s,0,t,\,;\,0,0,0)^{\text{\rm t}},\\
v_3(s,t)=&(0,0,1)^{\text{\rm t}}\otimes (s,t,0)^{\text{\rm t}}=(0,0,0,\,;\,0,0,0,\,;\,s,t,0)^{\text{\rm t}},\\
v_4(s,t)=&(0,s,t)^{\text{\rm t}}\otimes (1,0,0)^{\text{\rm t}}=(0,0,0,\,;\,s,0,0,\,;\,t,0,0)^{\text{\rm t}},\\
v_5(s,t)=&(s,0,t)^{\text{\rm t}}\otimes (0,1,0)^{\text{\rm t}}=(0,s,0,\,;\,0,0,0,\,;\,0,t,0)^{\text{\rm t}},\\
v_6(s,t)=&(s,t,0)^{\text{\rm t}}\otimes (0,0,1)^{\text{\rm t}}=(0,0,s,\,;\,0,0,t,\,;\,0,0,0)^{\text{\rm t}},\\
v_7(x_2,x_3)=&(1,x_2,x_3)^{\text{\rm t}}\otimes {\textstyle(1,\frac1{x_2},\frac1{x_3})^{\text{\rm t}}
=(1,\frac1{x_2},\frac1{x_3},\,;\,x_2,1,\frac{x_2}{x_3},\,;\,x_3,\frac{x_3}{x_2},1)^{\text{\rm t}}},
\end{aligned}
\end{equation}
where $s$ and $t$ are arbitrary complex numbers, and $x_2$ and $x_3$ are nonzero complex numbers.
Note that for $i=1,2,\dots,7$ and for any given pair $(s_i,t_i)$ of complex numbers with $s_7,t_7\neq 0$,
we can find pairs of complex numbers $(s_j,t_j)$
($1\le j\neq i \le 7)$ satisfying the following two conditions
\begin{equation}\label{eq:const1}
\dfrac 1{|s_7|^2}+|s_1|^2+|s_5|^2=\dfrac {|s_7|^2}{|t_7|^2}+|t_2|^2+|t_6|^2=|t_7|^2+|s_3|^2+|t_4|^2,
\end{equation}
and
\begin{equation}\label{eq:const2}
\dfrac 1{|t_7|^2}+|s_6|^2+|t_1|^2=\dfrac{|t_7|^2}{|s_7|^2}+|t_3|^2+|t_5|^2=|s_7|^2+|s_2|^2+|s_4|^2.
\end{equation}
For a $7$-tuple $\mathcal V=(v_1(s_1,t_1), v_2(s_2,t_2),\cdots,v_7(s_7,t_7))$
of the product vectors satisfying the conditions \eqref{eq:const1} and \eqref{eq:const2}, one can easily show that
\begin{equation}\label{eq:A1bc}
\dfrac 19 \sum_{i=1}^7\sum_{(\omega,\eta)\in\Omega\times \Omega}v_i(\omega s_i,\eta t_i)v_i(\omega s_i,\eta t_i)^*
=A[1,b(\mathcal V),c(\mathcal V)]
\end{equation}
where $b(\mathcal V)$ denotes the value in \eqref{eq:const2}, and $c(\mathcal V)$ denotes the value in \eqref{eq:const1}.
From the following inequalities
$$
\begin{gathered}
\left(\dfrac 1{|t_7|^2}+|s_6|^2+|t_1|^2\right)(|t_7|^2+|s_3|^2+|t_4|^2)\ge 1,\\
\left(\dfrac 1{|s_7|^2}+|s_1|^2+|s_5|^2\right)(|s_7|^2+|s_2|^2+|s_4|^2)\ge 1, \\
\left(\dfrac {|s_7|^2}{|t_7|^2}+|t_2|^2+|t_6|^2\right)\left(\dfrac{|t_7|^2}{|s_7|^2}+|t_3|^2+|t_5|^2\right)\ge 1,
\end{gathered}
$$
we know that $b(\mathcal V)c(\mathcal V)\ge 1$. Furthermore, we can show that
\begin{equation}\label{cond_bc}
b(\mathcal V)\ge 1,\quad c(\mathcal V)\ge 1.
\end{equation}
To see this, we assume that $b(\mathcal V)<1$ in \eqref{eq:const2}. Then it gives rise to the follwing inequalities:
\begin{equation}\label{contra}
\dfrac 1{|t_7|^2}<1,\quad \dfrac{|t_7|^2}{|s_7|^2}<1,\quad |s_7|^2<1.
\end{equation}
From the first two inequalities in \eqref{contra}, we abtain $\frac{1}{|s_7|^2}<1$.
This contradicts  the inequality $|s_7|^2<1$ in \eqref{contra}. Therefore, we conclude $b(\mathcal V)\ge 1$.
Similarly, we can show that $c(\mathcal V)\ge 1$.

For the equality in \eqref{cond_bc}, we have
$b(\mathcal V)=c(\mathcal V)=1$ holds if and only if the only nonzero product vector
in $7$-tuple $\mathcal V$ is $v_7(s_7,t_7)$ with $|s_7|=|t_7|=1$. In this case, the relation \eqref{eq:A1bc}
is reduced to the relation \eqref{eq:A111}. Note that $A[v_1]$ is on the boundary of the face
$\mathbb T[A[f]]$.

Now we proceed to show that $A[f]$ satisfies the condition (iv) of Theorem~\ref{the}.
We take arbitrary $z\in Q[A[1,b,c]]$ for fixed $b>1$ and $c>1$.
Then we may assume that $z$ is one of the vectors in \eqref{eq:vec_1bc}. For this product vector,
we can find $7$-tuple $\mathcal V$ satisfying the condition \eqref{eq:A1bc}.
Furthermore, we can choose $t_0>1$ satisfying the conditions
$$
(1-t_0)b(\mathcal V)+t_0 b >1\ \text{and}\  (1-t_0)c(\mathcal V)+t_0 c >1.
$$
Then we take vectors $v_1(\tilde s_1,\tilde t_1),\,v_2(\tilde s_2,\tilde t_2)$ and $v_3(\tilde s_3,\tilde t_3)$ in \eqref{eq:vec_1bc} as follows:
$$
\begin{gathered}
\tilde t_1=\tilde s_2=\tilde t_3=(1-t_0)b(\mathcal V)+t_0 b -1,\\
\tilde s_1=\tilde t_2=\tilde s_3=(1-t_0)c(\mathcal V)+t_0 c -1.
\end{gathered}
$$
Consequently, we have the relation
$$
A[1,b,c]=\dfrac 1{t_0} \left (A[1,1,1]+\sum_{i=1}^3 v_i(\tilde s_i,\tilde t_i)v_i(\tilde s_i,\tilde t_i)^* \right)
+\left(1-\dfrac 1{t_0}\right) A[1,b(\mathcal V),c(\mathcal V)].
$$
This shows that $A[f]$ satisfies the conditions of Theorem~\ref{the}.
We summarize as follows:
\begin{theorem}{Theorem}
We have the following:
\begin{itemize}
\item[{\rm (i)}]
$\mathbb V_1[A[e^1]]=\mathbb V_1\cap \mathbb T[A[e^1]]=\Phi(1)'$.
\item[{\rm (ii)}]
$\mathbb V_1[A[e^b]]=\mathbb V_1\cap \mathbb T[A[e^b]]$, for every nonnegative real number $b$ with $b\neq 1$.
\item[{\rm (iii)}]
$\mathbb V_1[A[f]]=\mathbb V_1\cap \mathbb T[A[f]]$.
\end{itemize}
\end{theorem}
\begin{proof}
It remains that $\mathbb V_1\cap \mathbb T[A[e^1]]=\Phi(1)'$.
 It is straightforwad to show that $zz^*\in \Phi(1)'$
for all product vector $z\in Q[A[e^1]]$. So we have the inclusion
$$
\mathbb V_1\cap \mathbb T[A[e^1]]\subset \Phi(1)'.
$$
For any product vector $z=(x_1,x_2,x_3)^{\text{\rm t}}\otimes (y_1,y_2,y_3)^{\text{\rm t}}\,$, we see that
$$
\begin{aligned}
zz^*\in \Phi(1)'
\Longleftrightarrow & |x_1\bar y_2-x_2 \bar y_1|^2+|x_1\bar y_3-x_3 \bar y_1|^2+|x_2\bar y_3-x_3 \bar y_2|^2=0\\
\Longleftrightarrow & (\bar x_1,\bar x_2,\bar x_3)^{\text{\rm t}}\,\,\text{is parallel to }\,(y_1,y_2,y_3)^{\text{\rm t}}\,\\
\Longleftrightarrow & z\in Q[A[e^1]].
\end{aligned}
$$
Therefore we can conclude that $\Phi(1)'$ coincides with the face $\mathbb V_1\cap \mathbb T[A[e^1]]$.
\end{proof}

In conclusion, we exhibited examples of separable states which are on the boundary of the convex cone $\mathbb V_1$ generated by all
separable states, but lie in the interior of the convex cone $\mathbb T$ of all PPT states. We also showed that they determine
maximal faces of the cone $\mathbb V_1$ whose interiors lie in the interior of the cone $\mathbb T$, and found all extreme rays in these
maximal faces. We note that there are recent remarkable progresses \cite{alfsen,alfsen_2} to understand the facial structures
of the convex cone $\mathbb V_1$. They are very useful when we consider the faces of the cone $\mathbb V_1$ which is
determined by separable states whose range dimensions are low. It is the hope of the authors that our examples motivate
general framework to deal with faces of the cone $\mathbb V_1$ determined by arbitrary separable states.


\begin{thebibliography}{999}

\bibitem{alfsen}
E. Alfsen and F. Shultz,
\it Unique decompositions, faces, and automorphisms of separable states,
\rm J. Math. Phys. \bf 51 \rm (2010), 052201.

\bibitem{alfsen_2}
E. Alfsen and F. Shultz,
\it Finding decompositions of a class of separable states,
\rm preprint,
arXiv:1202.3673.

\bibitem{BZ}
I. Bengtsson and K. \.Zyczkowski,
Geometry of Quantum States: An Introduction to Quantum Entanglement,
Cambridge University Press, 2006.

\bibitem{cho-kye-lee}
S.-J. Cho, S.-H. Kye and S. G. Lee,
\it Generalized Choi maps in $3$-dimensional matrix algebras,
\rm Linear Alg. Appl. \bf 171 \rm (1992), 213--224.

\bibitem{choi_kye}
H.-S. Choi and S.-H. Kye,
\it Facial structures for separable states,
\rm J. Korean Math. Soc. {\bf 49} (2012), 623--639.

\bibitem{choi-ppt}
M.-D. Choi,
\it Positive linear maps,
\rm Operator Algebras and Applications (Kingston, 1980), pp. 583--590,
Proc. Sympos. Pure Math. Vol 38. Part 2, Amer. Math. Soc., 1982.

\bibitem{choi-lam}  M.-D. Choi and T.-T. Lam,
\it Extremal positive semidefinite forms,
\rm Math. Ann. \bf 231 \rm (1977), 1--18.


\bibitem{cs}
D. Chru\'{s}ci\'{n}ski and G. Sarbicki,
\it Optimal entanglement witnesses for two qutrits,
\rm preprint,
arXiv:1108.0513.

\bibitem{cw}
D. Chru\'{s}ci\'{n}ski and F. A. Wudarski,
\it Geometry of entanglement witnesses for two qutrits,
\rm Open Syst. Inf. Dyn. {\bf 18} (2011), 375--387.

\bibitem{eom-kye}
M.-H. Eom and S.-H. Kye,
\it Duality for positive linear maps in matrix algebras,
\rm Math. Scand. \bf 86 \rm (2000), 130--142.

\bibitem{ha_kye_04}
K.-C. Ha and S.-H. Kye,
\it Construction of entangled states with positive partial
transposes based on indecomposable positive linear maps,
\rm Phys. Lett. A \bf 325 \rm (2004), 315--323.

\bibitem{ha+kye_indec-witness}
K.-C. Ha and S.-H. Kye,
\it One parameter family of indecomposable optimal entanglement witnesses arising from generalized Choi maps,
\rm Phys. Rev. A, \bf 84 \rm (2011), 024302.

\bibitem{ha+kye_exposed}
K.-C. Ha and S.-H. Kye,
\it Entanglement witnesses arising from exposed positive linear maps,
\rm Open Syst. Inf. Dyn. \bf 18 \rm (2011), 323--337.

\bibitem{ha+kye}
K.-C. Ha, S.-H. Kye and Y. S. Park,
\it Entanglements with positive partial transposes arising from indecomposable positive linear maps,
\rm Phys. Lett. A {\bf 313} (2003),  163--174.

\bibitem{horo-1}
M. Horodecki, P. Horodecki and R. Horodecki,
\it Separability of mixed states: necessary and sufficient conditions,
\rm Phys. Lett. A \bf 223 \rm (1996), 1--8.

\bibitem{mpr_horo}
M. Horodecki, P. Horodecki and R. Horodecki, \it Mixed-state
entanglement and distillation: is there a ``bound'' entanglement in
nature?, \rm Phys. Rev. Lett. \bf 80 \rm (1998), 5239--5242.


\bibitem{hugh}
L. P. Hughston, R. Jozsa and W. K. Wootters,
\it A complete classification of quantum ensembles having a given density matrix,
\rm Phys. Lett. A \bf 183 \rm (1993), 14--18.

\bibitem{kye-canad}
S.-H. Kye,
\it Facial structures for positive linear maps between matrix algebras,
\rm Canad. Math. Bull. \bf 39 \rm (1996), 74--82.

\bibitem{kye_ritsu}
S.-H. Kye,
\it Facial structures for various notions of positivity and applications to the theory of entanglement,
\rm preprint, arXiv:1202.4255.

\bibitem{kye_osaka}
S.-H. Kye and H. Osaka,
\it Classification of bi-qutrit PPT entangled edge states by their ranks,
\rm preprint,
arXiv:1202.1699.

\bibitem{peres}
A. Peres,
\it Separability Criterion for Density Matrices,
\rm Phys. Rev. Lett. \bf 77 \rm (1996), 1413--1415.



\bibitem{stormer82}
E. St\o rmer,
\it Decomposable positive maps on $C^*$-algebras,
\rm Proc. Amer. Math. Soc. \bf 86 \rm (1982), 402--404.




\end{thebibliography}
\end{document}